\documentclass[usenatbib,fleqn]{mnras}
\usepackage{graphicx}
\usepackage{amssymb}
\usepackage{amsmath}

\newcommand{\lcur}{\mathcal{L}}
\newcommand{\Rcur}{\mathcal{R}}

\newcommand{\tm}{t_{\rm m}}
\newcommand{\Nobs}{N_{\rm obs}}
\newcommand{\Nmod}{N_{\rm mod}}
\newcommand{\Mr}{M_{\rm r}}

\usepackage[usenames,dvipsnames]{xcolor}

\title[DTD of BNSM from Host Galaxy Properties]{Constraining Delay Time Distribution of Binary Neutron Star Mergers from Host Galaxy Properties}

\author[K.S. McCarthy, Z. Zheng, and E. Ramirez-Ruiz]{
Kevin S. McCarthy$^{1}$\thanks{E-mail: kevin.mccarthy@utah.edu},
Zheng Zheng$^{1}$\thanks{E-mail: zhengzheng@astro.utah.edu},
and Enrico Ramirez-Ruiz$^{2,3}$\thanks{E-mail: enrico@ucolick.org}\\
\\
$^{1}$ Department of Physics and Astronomy, University of Utah, Salt Lake City, UT 84112, USA\\
$^{2}$ Department of Astronomy and Astrophysics, University of California, Santa Cruz, CA 95064, USA\\
$^3$DARK, Niels Bohr Institute, University of Copenhagen, Blegdamsvej 17, 2100 Copenhagen, Denmark
}

\begin{document}
\label{firstpage} \pagerange{\pageref{firstpage}--\pageref{lastpage}}
\maketitle
\begin{abstract}
Gravitational wave (GW) observatories are discovering binary neutron star mergers (BNSMs), and in at least one event we were able to track it down in multiple wavelengths of light, which allowed us to identify the host galaxy. 
Using a catalogue of local galaxies with inferred star formation histories and adopting a BNSM delay time distribution (DTD) model, we investigate the dependence of BNSM rate on an array of galaxy properties. Compared to the intrinsic property distribution of galaxies, that of BNSM host galaxies is skewed toward galaxies with redder colour, lower specific star formation rate, higher luminosity, and higher stellar mass, reflecting the tendency of higher BNSM rates in more massive galaxies. We introduce a formalism to efficiently make forecast on using host galaxy properties to constrain DTD models. We find comparable constraints from the dependence of BNSM occurrence distribution on galaxy colour, specific star formation rate, and stellar mass, all better than those from dependence on $r$-band luminosity. The tightest constraints come from using individual star formation histories of host galaxies, which reduces the uncertainties on DTD parameters by a factor of three or more. Substantially different DTD models can be differentiated with about 10 BNSM detections. To constrain DTD parameters at 10\% precision level requires about one hundred detections, achievable with GW observations on a decade time scale.
\end{abstract}

\section{Introduction}\label{sec:intro}
The dawn of multi-messenger astronomy began with the observation of a binary neutron star merger (BNSM; \citealt{Abbott17mm}). Originating in the galaxy NGC 4993 \citep{Levan17} located at a distance $41\pm 3.1$ Mpc \citep{Hjorth17}, two neutron stars in orbit about each other merged together, emitting waves not only across the electromagnetic (EM) spectrum but also in spacetime. Gravitational waves (GW) from this event (GW170817; \citealt{Abbott17}) were detected by the LIGO-Virgo Collaboration detector network. A couple seconds after the GW signal, the Fermi Gamma-ray Burst Monitor detected a short gamma-ray burst (GRB) (GRB 170817A; \citealt{Goldstein17}). The chirp mass and presence of a short GRB indicated that this event was from a BNSM, and an extensive optical campaign was launched to search for the EM counterpart. In about $11$ hours, the One-Meter Two-Hemispheres Collaboration discovered a transient and fading optical source with the Swope Telescope in Chile (SSS17a; \citealt{Coulter17}) coincident with GW170817. The observation of this BNSM event, in many aspects, marked a transition in our knowledge from being purely theoretical to, now, empirical.

It has been known since the detection of the orbital decay of a binary pulsar \citep{Hulse75} that these systems are radiating GW, implicit according to general relativity (GR). What is not so evident is how these systems form and what happens in the final moments of their merger. It had been proposed that these mergers should be extremely luminous, releasing high energy photons in the form of short GRBs \citep{Lee2007,Berger10,Berger13,Fong15}, activating the rapid neutron capture process ($r$-process; \citealt{Symbalisty82,Freiburghaus99}), and forming kilonova events \citep{Eichler89,Li98,Metzger2010,Roberts2011,Kasen2017}. Such predictions are confirmed by the detection of EM counterparts associated with GW170817 \citep{Kilpatrick2017, Murguia-Berthier2017, Evans17, Tanvir17, Hotokezaka18, Wu19}. What is not yet well understood is whether BNSMs can account for the abundance of $r$-process elements observed in the Milky Way \citep[e.g.][]{Macias2018} and whether they are the progenitors of all observed short GRBs \citep[e.g.][]{Behroozi2014}. This requires a deep understanding of the BNSM merger channel, which will in turn elucidate how often these type of events occur. Conversely, observational constraints on the BNSM GW event rate will uncover the likely distribution of their merger times and thus the important physical mechanisms in play \citep[e.g.][]{Kelley2010}.

The delay-time distribution (DTD) of BNSMs is a short hand description that encapsulates all the physical mechanisms from the time of formation of stellar mass to the moment of the final merger event \citep[e.g.][]{Vigna2018}, including the main-sequence lifetime of the progenitor stars, their post main-sequence evolution, and various phases of binary evolution \citep[such as supernova explosion and the common-envelope phase;][]{Fragos2019}. DTD is likely dominated by the in-spiral time caused by GW radiation. The delay-time scale for a binary system is predicted by GR as $t\propto a^4(1-e^2)^{7/2}$, with $a$ the initial semi-major axis and $e$ the eccentricity of the system. For circular orbits ($e=0$), the distribution of $a$ is usually characterised to follow a power-law form, $dN/da \propto a^{-p}$, which implies the DTD $dN/dt \propto t^n$ with $n=-(p+3)/4$. If $a$ follows a uniform distribution in log-space (i.e. $p=1$), the DTD then has a power-law index $n=-1$ \citep{Piran92,Beniamini19}.

This canonical, in-spiral dominated, DTD with $n=-1$ is supported by evolutionary modelling of the BNSM \citep{Dominik12, Belczynski18}, as well as the inference of merger times in observed Galactic binary neutron star systems \citep{Beniamini19}. However, it is argued that $n=-1$ might not be steep enough to produce the observed abundances of $r$-process elements (e.g. Europium) in the Milky Way \citep{Cote17,Simonetti19,Beniamini19}, which might require shorter merger times 
or an improvement in our current understanding of turbulent mixing in the early Milky Way \citep{Shen2015,Naiman2018}. In the case of GW170817,  \cite{Belczynski18} find that the canonical DTD has too short merger times to make GW170817 a typical BNSM event, since NGC 4993 is a galaxy dominated by an old stellar population \citep{Blanchard17}. \citet{Fong17} also find that NGC 4993 is atypical in many ways to the observed host galaxies of short GRB events, suggesting the possibility that GW170817 may not be representative of BNSM events.

More detections of GW events from BNSM are thus needed to have meaningful constraints on the corresponding DTD. Future constraints have been investigated based on distribution of stellar mass of BNSM host galaxies \citep{Safarzadeh19a}, redshift distribution of BNSM events \citep{Safarzadeh19b}, and star formation history (SFH) of individual host galaxies \citep{Safarzadeh19c}. \citet{Adhikari20} study the properties of host galaxies of BNSM events based on a Universe Machine simulation of galaxy evolution and discuss the constraints on the DTD models. \citet{Artale19} and \citet{Artale20} combine BNSM models from population synthesis with galaxy catalogues in hydrodynamic galaxy formation simulations to study the correlation of BNSM rate with galaxy properties. 

In this paper, based on a galaxy catalogue in the local universe, we investigate the connection between the DTD and various galaxy properties, formulate an efficient method to forecast the DTD constraints from distributions of BNSM host galaxy properties and from their individual SFH, and present the forecasts on DTD constraints for future GW observations, which will also benefit the efforts of localising the EM counterparts and searching for the host galaxies. In Section~\ref{sec:data}, we introduce the galaxy catalogue used in the study. In Section~\ref{sec:method}, we introduce the methodology and the forecast formalism. The main results are presented in Section~\ref{sec:results}. After a discussion in Section~\ref{sec:dis}, we summarise and conclude the work in Section~\ref{sec:sum}.

\section{Data}\label{sec:data}
Our investigation makes use of the main galaxy sample from the Sloan Digital Sky Survey (SDSS; \citealt{York00}) Data Release 7 (DR7; \citealt{Abazajian09}). We include in the study the following properties of galaxies, luminosity (absolute magnitude), colour, SFH, stellar mass, and specific star formation rate (sSFR).

The $r$-band absolute magnitude $\Mr^{0.1}-5\log h$ and $(g-r)^{0.1}$ colour are from the New York University value-added catalogue (NYU VAGC; \citealt{Blanton05, Padmanabhan08}), which have been K+E corrected to redshift $z=0.1$ (thus the superscript) according to WMAP3 spatially-flat cosmology \citep{Spergel07} with $\Omega_{\rm m}=0.238$ and $H_0=100h{\rm \, km\, s^{-1} Mpc^{-1}}$ with $h=0.732$. For simplicity and without confusion, we remove the superscript $0.1$ hereafter.

The SFH of each SDSS galaxy is pulled from the versatile spectral analysis (VESPA; \citealt{Tojeiro07,Tojeiro09}) database. SFH is derived based on the stellar population synthesis model of \citeauthor{Bruzual03} (\citeyear{Bruzual03}; BC03) with uniform dust extinction. We use the data with the highest temporal resolution -- for each galaxy, star formation rate (SFR) is stored in 16 logarithmically-spaced lookback time bins from 0.02 to 14 Gyr (see fig.1 of \citealt{Tojeiro09}), with the zero point of lookback time determined by its redshift. VESPA employs the WMAP5 cosmology \citep{Komatsu09} with  $\Omega_{\rm m}=0.273$ and $h=0.705$ to shift the galaxy spectra to rest-frame.

The stellar mass ($M_*$; \citealt{Kauffmann03,Salim07}) and sSFR (defined as SFR/$M_*$; \citealt{Brinchmann04}) come from the Max Planck for Astrophysics and Johns Hopkins University value-added catalogue (MPA-JHU VAGC; \citealt{Tremonti04}), both estimated for $z=0.1$. Stellar mass is derived through fits to a large grid of SFHs using the BC03 model and sSFR is determined through emission line features and/or the 4000\AA-break. While stellar masses employ photometry calculated under WMAP3 cosmology, the sSFR calculation assumes a cosmology with $\Omega_{\rm m}= 0.3$ and $h= 0.7$. 

Since we focus on local galaxies ($z\sim 0.1$), the differences in cosmology used in the DR7 photometry, VESPA, and MPA/JHU analyses lead to no significant consequences at all. With all the properties, we end up with $\sim 515$K galaxies.
Further inspection of each galaxy's SFH shows that some have exorbitant stellar mass formed in a particular lookback time bin relative to the general population, and we find that their spectra have been contaminated by spurious signal, i.e. cosmic rays. We apply a $6\sigma$ clip according to the $\log({\rm SFR})$ distribution in particular temporal bins and also remove those in the noisy tail distribution. In the end, we have a galaxy catalogue composed of $\sim 501$K galaxies, with properties $\Mr$, $g-r$, SFH, $M_*$, and sSFR, allowing accurate characterisations of distributions of galaxy properties to be used in our investigations. 

Specifically, galaxies in our catalogue are selected based on the following luminosity and colour cuts, $\Mr-5\log h$ in the range (-22.0, -16.5) and $g-r$ in the range (0.0, 1.2). Our calculations effectively use a volume-limited sample of galaxies (see Section~\ref{sec:method}). Given the exponential cutoff in galaxy luminosity function at the high luminosity end and the power-law behaviour at the low luminosity end, in computing the BNSM rate, we mainly miss the contribution from galaxies dimmer than $\Mr-5\log h=-16.5$ mag. With the selection, in terms of stellar mass, the sample mainly becomes incomplete at $\lesssim 10^9M_\odot$ (Section~\ref{sec:prop}). The incompleteness in the low-luminosity or low-stellar-mass galaxies does not affect our results. First, the contribution to the BNSM rate from galaxies below the luminosity cut is small, estimated to be about 4\% even for the most extreme model we consider (see Section~\ref{sec:prop}). Second, the analysis can be thought as to use BNSM events detected in galaxies satisfying the above luminosity and colour cuts.


\section{Method}\label{sec:method}

\subsection{BNSM Rate Calculation and DTD Parameterisation}

In our study, we group galaxies according to their properties. We investigate the dependences of BNSM rate on various galaxy properties and how such dependences help constrain the DTD. The ultimate limit is to use SFH information of each individual host galaxies.

For a galaxy with SFH given by the time-dependent SFR, the expected BNSM rate reads \citep[e.g.][]{Zheng07}
\begin{equation}
    \Rcur=C\int_0^{t_{\rm max}} {\rm SFR(\tau)} P(\tau) {\rm d}\tau,
    \label{eq:rate}
\end{equation}
where $P$ is the DTD function. The integral variable is put in terms of the lookback time $\tau$ with respect to that at the redshift of the galaxy, following the way how the SFH is stored in the data. Subsequently $t_{\rm max}$ is the age $t_0$ of the universe minus the lookback time to the galaxy redshift, and for local galaxies $t_{\rm max}\sim t_0$. 
The constant $C$ relies on details of the formation and evolution of binary neutron star systems, which can be determined for a given model of the stellar and binary populations. Since our study uses the relative distribution of BNSM rate as a function of galaxy properties, this constant plays no role.

When galaxies are grouped by a property, we compute the mean BNSM rate based on the average SFR within each bin of the property. To account for the observational limit of galaxies, we weigh each galaxy by $1/V_{\rm max}$, where $V_{\rm max}$ is the maximum volume that the galaxy can be observed given the limiting magnitude of the survey. That is, we use the number density of galaxies in each property bin, $n_{\rm g}=\sum_i 1/V_{{\rm max},i}$, where $i$ denotes the $i$-th galaxy in the bin. Therefore, our results are effectively for a volume-limited sample of galaxies.

We parameterise the DTD function as (e.g. \citealt{Safarzadeh19a})
\begin{equation}
    P(\tau; n,t_{\rm m}) \propto \left\{
    \begin{array}{ll}
    0,      & \tau<t_{\rm m},\\
    \tau^n, & \tau\geq t_{\rm m}.
    \end{array}
    \right.
\end{equation}
That is, the distribution follows a power-law with index $n$, which has a cutoff at $t_{\rm m}$. This minimum delay time $\tm$ encodes information about the formation and evolution of the binary system, including time from star formation to supernova explosion and the distribution of binary orbits.

\subsection{Likelihood Calculation and Forecast Formalism}
\label{sec:likelihood}

To perform forecast on using BNSM GW events with associated host galaxy properties to constrain DTD, we employ the likelihood analysis.

For the dependence of BNSM rate on a certain galaxy property (e.g. stellar mass), following \citet{Gould95}, we divide our galaxy sample into small bins of the property and in each bin the BNSM occurrence is assumed to follow Poisson distribution. If during an observation period we observe $k_i$ events in the $i$-th bin, for a DTD model that predict a mean number of $\lambda_i$ events in the bin, the total likelihood is then
\begin{equation}
    \lcur = \prod_i 
    \frac{\lambda_i^{k_i} {\rm e}^{-\lambda_i}}{k_i!}, 
\label{eq:likePDF}
\end{equation}
where the multiplication goes through all the property bins. We will work in the regime that the bins are sufficiently small such that $k_i$ is either 0 or 1 (i.e. $k_i!=1$). In terms of the log-likelihood, we have
\begin{equation}
     \ln \lcur = 
     \sum_i k_i \ln\lambda_i - \sum_i \lambda_i - \sum_i \ln k_i!
     = \sum_i k_i \ln\lambda_i - \Nmod,
     \label{eq:logL}
\end{equation}
where $\Nmod=\sum_i \lambda_i$ is the total number of events predicted by the model. 

In order to do the forecast, we need to assume a underlying truth model, which generates the observation. We use `*' to label quantities from the truth model and denote the mean number of events in the $i$-th bin as $\lambda^*_i$ and the total predicted number as $\Nobs=\sum_i \lambda^*_i$ for the truth model. The series of $k_i$ in equation~(\ref{eq:logL}) form a realisation of the truth model. For the given realisation the likelihood function we need to evaluate is then
\begin{equation}
\Delta\ln\lcur = \ln \lcur - \ln \lcur^*  =   
\sum_i k_i \ln\frac{\lambda_i}{\lambda^*_i} - \Nmod + \Nobs.
\label{eq:diff_logL}
\end{equation}
With this equation, the evaluation of the likelihood for any model can be made for a given observation (i.e. the $k_i$ series). A large number of realisations of observation with different series of $k_i$ generated by the truth model can be performed. There are variations among different realisations and an average over realisations can be used for forecasting the DTD parameter constraints \citep[e.g.][]{Safarzadeh19b}.

Here we avoid performing the realisations by considering the ensemble average of equation~(\ref{eq:diff_logL}),
\begin{equation}
     \langle \Delta\ln\lcur \rangle = 
\sum_i \lambda^*_i \ln\frac{\lambda_i}{\lambda^*_i} - \Nmod + \Nobs,
     \label{eq:avg_diff_logL}
\end{equation}
where the ensemble average of the number of observed events in the $i$-th property bin, $\langle k_i \rangle$, is just the mean number $\lambda_i^*$ from the truth model. With the ensemble average likelihood, we effectively have an average realisation that can be efficiently evaluated as shown below.

The mean number $\lambda_i$ for a model is calculated from equation~(\ref{eq:rate}) and galaxy property distribution. In fact, we can compute the probability density distribution $p(x)$ as a function of galaxy property $x$. In the $i$-th bin with property $x_i$ and bin width $\Delta x_i$, 
\begin{equation}
p(x_i)\Delta x_i = \frac{ n_{{\rm g},i}\Rcur_i }{\sum_j n_{{\rm g},j}\Rcur_j},
\label{eq:prob}
\end{equation}
where $n_{{\rm g},i}$ is the number density of galaxies in the bin and $\Rcur_i$ the BNSM rate from the mean SFH of those galaxies in the bin. We note that the bin width $\Delta x_i$ should be understood as multi-dimensional, e.g. the size of the colour-magnitude bin if we are to consider the dependence of the BNSM distribution on the host galaxy's colour and magnitude. Clearly $p(x_i)$ is independent of the constant $C$ in equation~(\ref{eq:rate}). As this probability distribution is normalised by definition, $\sum_i p(x_i) \Delta x_i=1$, we can write $\lambda_i=\Nmod\, p(x_i) \Delta x_i$ and similarly $\lambda_i^*=\Nobs\, p^*(x_i) \Delta x_i$.  Equation~(\ref{eq:avg_diff_logL}) then becomes
\begin{flalign}
\langle \Delta \ln \lcur \rangle 
     & =  \Nobs \sum_i p^*(x_i) \ln \frac{p(x_i)}{p^*(x_i)} \Delta x_i \nonumber \\
     & \,\,\,\,\,\, + \Nobs \ln \frac{\Nmod}{\Nobs} -\Nmod+\Nobs \nonumber \\ 
     & =  N_{\rm obs} \int p^*(x) \ln \frac{p(x)}{p^*(x)} {\rm d} x \nonumber\\
     & \,\,\,\,\,\, + \Nobs \ln \frac{\Nmod}{\Nobs} -\Nmod+\Nobs,
     \label{eq:fullLike}
\end{flalign}
where in the last step we have taken the limit $\Delta x_i\rightarrow 0$. 

Note that, in the analysis, the dependence on galaxy property lies in $p$ (as well as $p^*$), which is determined by the number density of galaxies and the mean SFR in each bin of the galaxy property in consideration [equations~(\ref{eq:rate}) and (\ref{eq:prob})].

As we focus on studying the BNSM distribution as a function of a given galaxy property, we can always normalise any model to have $N_{\rm mod}=N_{\rm obs}$. The function to evaluate then becomes
\begin{equation}
    \langle \Delta \ln \lcur\rangle =  N_{\rm obs} \int p^* \ln\frac{p}{p^*} {\rm d}x.
    \label{eq:DeltaL}
\end{equation}
Interestingly but not surprisingly, the likelihood ratio is related to the relative entropy of two distributions \citep{Kullback51}. Given a truth model, for each model to be evaluated we only need to calculate the integral on the right-hand side once. The nice and simple scaling relation with $N_{\rm obs}$ makes it easy to investigate the dependence of parameter constraints on the number of observations.

For constraints making use of SFH of individual galaxies, it is easy to show that equation~(\ref{eq:DeltaL}) takes the form
\begin{equation}
    \langle \Delta \ln \lcur\rangle = N_{\rm obs} \sum_i p_i^*\ln\frac{p_i}{p_i^*},
    \label{eq:DeltaL_perGAL}
\end{equation}
where $i$ denotes the $i$-th galaxy. The probability $p_i$ can be calculated as the rate $\Rcur_i$ from equation~(\ref{eq:rate}) expected for the $i$-th galaxy divided by the total rate from all galaxies in consideration, $p_i=\Rcur_i/\sum_j \Rcur_j$. As mentioned before, the rate is weighted by $1/V_{\rm max}$ for each galaxy as we consider an effectively volume-limited sample of galaxies.


We could continue to compute the second derivatives of equation~(\ref{eq:DeltaL}) or (\ref{eq:DeltaL_perGAL}) with respect to model parameters and perform Fisher matrix analysis \citep[e.g.][]{Tegmark97} to investigate the constraints. However, given that we only have two model parameters, we will evaluate the model likelihood on a grid of parameters to obtain an accurate description of the likelihood surface.

\section{Results}
\label{sec:results}

With the SFH information of the sample of SDSS galaxies, we first present the dependence of the occurrence distribution of BNSM events on galaxy properties for a set of DTD models. Then based on the formalism developed in Section~\ref{sec:likelihood} we make forecasts on constraining the DTD distribution with GW observations of BNSM events.

We choose three representative DTD models to illustrate the results, corresponding to a `Fast', a `Canonical', and a `Slow' merging channel, respectively:
\begin{itemize}
    \item  The `Fast' model has a steep slope ($n=-1.5$) and a short minimum delay time ($\tm=0.01$ Gyr), which is motivated by the requirement to have prompt injection of $r$-process material in the early evolution of the Milky Way (see Section~\ref{sec:intro}). \\
    \item The `Canonical' model represents the canonical, in-spiral dominated DTD, with $n=-1.1$ and $t_m=0.035$ Gyr. The power-law index comes from the constraints with the inferred DTD of Galactic binary neutron stars \citep{Beniamini19}. \\
    \item The `Slow' model, with $n=-0.5$ and $\tm=1$ Gyr, tends to increase the number of events in galaxies of old stellar populations, as hinted by the case of GW170817 \citep[e.g.][]{Blanchard17,Belczynski18}.\\
\end{itemize}
When presenting the forecasts on DTD parameter constraints, we consider three cases, with each of the above three models adopted as the truth model.

\subsection{Dependence of BNSM Occurrence on Galaxy Properties}
\label{sec:prop}

We start by studying the distribution of BNSM events as a function of both galaxy colour and luminosity, i.e. in the colour-magnitude diagram (CMD). Then we investigate the dependence on galaxy colour, luminosity, stellar mass, and sSFR, respectively. All calculations are based on equation~(\ref{eq:prob}).

\begin{figure*}
\includegraphics[width=\textwidth]{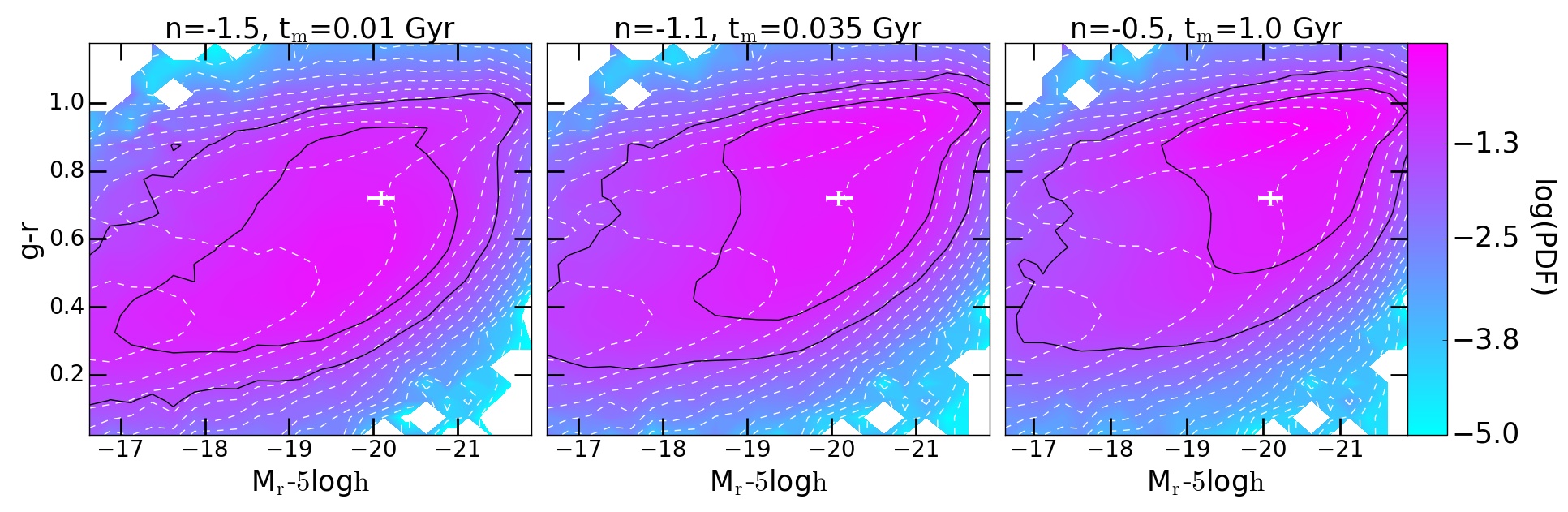}
\caption{
Dependence of occurrence probability distribution of BNSM events on galaxy colour ($g-r$) and luminosity ($\Mr-5\log h$). The calculation is done for a volume-limited sample of local galaxies, and the total probability is normalised to be unity over the range of colour and luminosity shown in each panel. The left, middle, and right panels are from the `Fast', `Canonical', and 'Slow' DTD model, respectively, with model parameters ($n$, $\tm$) labelled at the top of each panel. In each panel, the solid and dashed contours represent the 68.3\% ($1\sigma$) and 95.4\% ($2\sigma$) range of the distribution around the peak. The cross represents the colour and magnitude (with error bars) of NGC 4993, host galaxy of the BNSM event associated with GW170817.
}
\label{fig:PDFcmd}
\end{figure*}

In Fig.~\ref{fig:PDFcmd}, the probability distribution function (PDF) of BNSM rate is shown in the $\Mr$--$(g-r)$ plane of galaxies for each of the three selected DTD models. We consider all galaxies with $\Mr-5\log h$ in the magnitude range (-22.0, -16.5) and $g-r$ colour in the magnitude range (0.0, 1.2). The probability calculated according to equation~(\ref{eq:prob}) is essentially the usual galaxy CMD (i.e. galaxy number density distribution) convolved with the BNSM rate as a function of galaxy colour and luminosity. The two solid contours in each panel enclose the 68.3\% and 95.4\% of the BNSM rate distribution around the maximum, respectively. As a comparison, the dashed contours show the distribution of galaxy number density, where the blue cloud, the green valley, and the red sequence can be identified.

For the probability distribution of the `Fast' DTD model (left panel), the central 68.3\% distribution encloses the blue cloud galaxies at low luminosity and the red sequence galaxies up to $\approx L^*$ ($\Mr^*-5\log h=-20.44$ mag; \citealt{Blanton03}), as well as the green valley galaxies in between them. As the model prefers young stellar populations, redder galaxies (e.g. those toward the luminous end of the red sequence) do not contribute much.
Toward the blue and low-luminosity corner, the low stellar masses and thus low BNSM rate per galaxy lead to a decreasing contribution from these galaxies.

For the `Slow' DTD model (right panel), the central 68.3\% of the distribution includes red sequence galaxies more luminous than -18.5 mag (about 0.17$L^*$), the luminous tail ($\Mr-5\log h<-19.2$ mag) of the blue cloud galaxies, and the green valley galaxies in between them. The overall shift toward redder galaxies in comparison to the left panel is a consequence that the model favours old stellar populations.

The distribution from the the `Canonical' DTD model (middle panel) is in between the two above cases. While we still have red sequence galaxies similar to the right panel, the distribution extends to lower luminosity in the blue cloud, across the green valley.

The cross in each panel marks the colour and magnitude of NGC 4993, the host galaxy of GW170817, based on photometry from \cite{Blanchard17} and distance estimate from \cite{Hjorth17}. For consistency with the galaxy sample we use, we have K-corrected the photometry to $z=0.1$ and converted the magnitude to $\Mr-5\log h$. The colour and magnitude of this galaxy fall into the 68.3\% range of the distribution implied by each of the three DTD models considered here. Clearly more BNSM detections and observations of host galaxies are necessary to probe the distribution in the CMD and constrain the DTD model.

\begin{figure*}
\includegraphics[width=\textwidth]{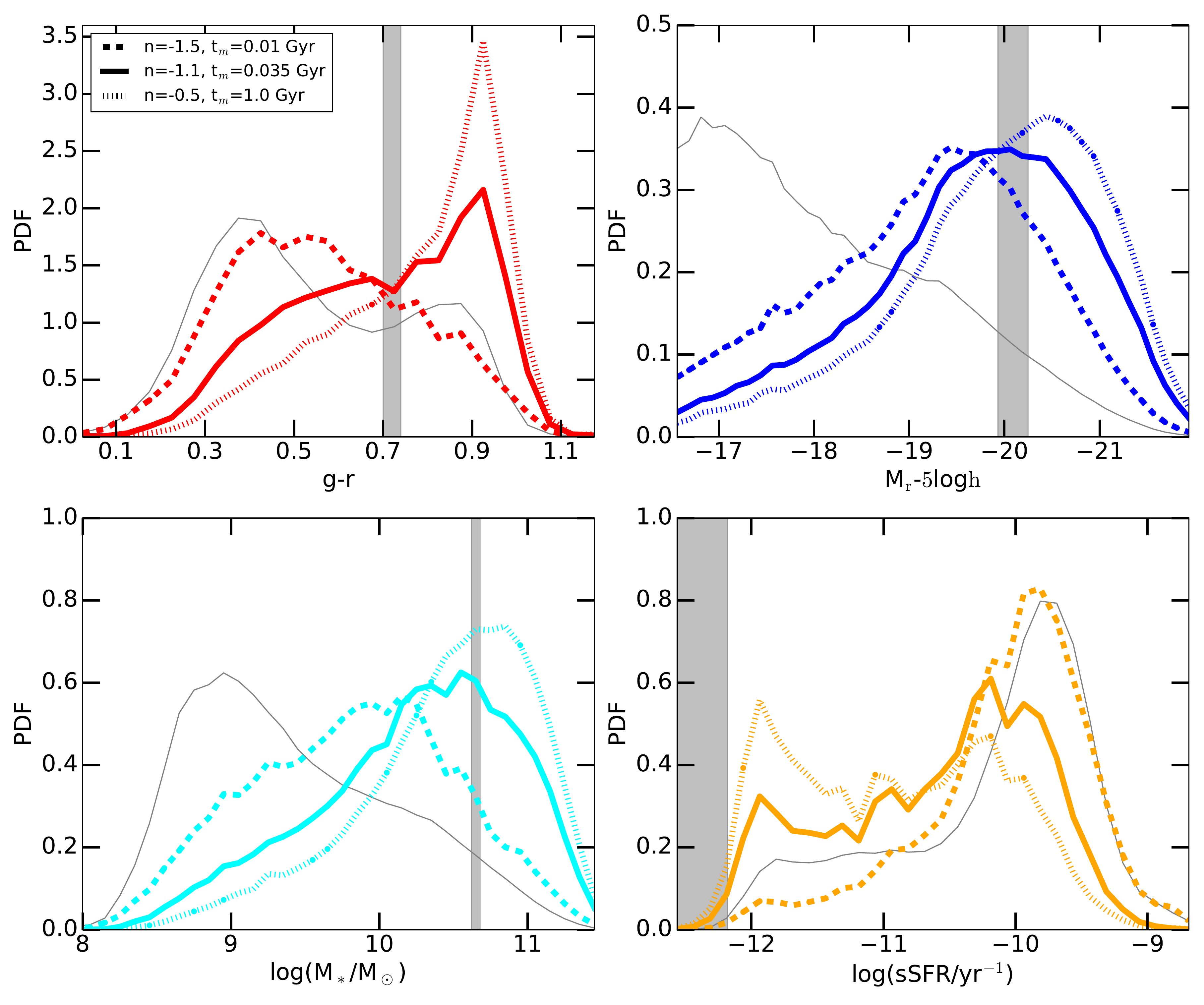}
\caption{
Dependence of occurrence probability distribution of BNSM events on galaxy colour (top-left), luminosity (top-right), stellar mass (bottom-left), and sSFR (bottom-right).
The calculation is done for a volume-limited sample of local galaxies. The dashed, solid, and dotted curves are for the `Fast', `Canonical', and 'Slow' DTD model, respectively. In each panel, the thin solid curve shows the intrinsic distribution of the galaxy property for local galaxies. The vertical band represents the range of the observed property of NGC 4993, host galaxy of the BNSM event associated with GW170817.
}
\label{fig:PDFmlt}
\end{figure*}

Next we turn to the dependence of probability distribution of BNSM events on each of the colour, luminosity, stellar mass, and sSFR, as shown in Fig.~\ref{fig:PDFmlt}. These four properties are broadly correlated, in the sense that on average redder galaxies are more luminous, higher in stellar mass, and lower in sSFR. Therefore the distributions shown in the four panels share similar trends. The distribution from the `Fast' DTD model (thick dashed), which favours younger stellar populations, peaks at bluer colour, lower luminosity, lower stellar mass, or higher sSFR than that from the `Slow' DTD model (thin dotted). The distribution from the `Canonical' model (thick solid) lies in between the above two cases. 

The thin solid curve in each panel of Fig.~\ref{fig:PDFmlt} shows the intrinsic distribution of galaxy property of the underlying galaxy sample we use, i.e. the distribution of galaxy number density. The BNSM host galaxy distribution is simply this galaxy property distribution modified by the property-dependent BNSM rate. In the top-left panel, we see the bimodal colour distribution of galaxies. On average, the BNSM rates [equation~(\ref{eq:rate})] are higher in redder galaxies, as they tend to be more massive (and thus on average higher SFR over the history). This gives higher weights to redder galaxies. As a consequence, the distribution of colour of host galaxies skews toward red colour and the original bimodal feature is smeared out. The case with the sSFR (lower-right panel) is similar.

The thin solid curve in the top-right panel Fig.~\ref{fig:PDFmlt} shows the intrinsic luminosity distribution of galaxies, which is proportional to the luminosity function. While there are a larger number of faint galaxies, their lower masses (thus on average lower SFR over the history) lead to lower contribution to BNSM rates. Our galaxy sample includes galaxies more luminous than $\Mr-5\log h = -16.5$ mag, and even with the most conservative estimate from the `Fast' DTD model, BNSMs from galaxies fainter than this limit only contribute $\sim$4\% of  events. The luminosity distribution of BNSM host galaxies tend to peak around $-20\pm 0.5$ mag. The situation with the stellar mass (bottom-left panel) is similar. Note that the galaxy sample we use is complete for galaxies more luminous than -16.5 mag, which is not complete in stellar mass at the low mass end. The scatter between luminosity and stellar mass causes the soft cutoff (around $10^9M_\odot$) in the low-mass end of the stellar mass distribution (thin curve in the bottom-left panel).

The vertical band in each panel of Fig.~\ref{fig:PDFmlt} indicates the property of the host galaxy of GW170817 \citep{Blanchard17}. The colour, magnitude, or stellar mass appears to be around the middle of the corresponding host galaxy distribution. So in terms of these three properties, the host galaxy of GW170817 is not atypical. However, the sSFR of this host galaxy appears to be at the very tail of the distribution, making it atypical in this regard.

As a whole, the above results show how the occurrence probability of BNSM events depends on galaxy properties and the DTD models. The three DTD models we present likely cover the range of models. Based on Fig.~\ref{fig:PDFcmd}, the most likely host galaxies of BNSM events (in the sense of the 68.3\% range of the distribution) lie within a diagonal band in the CMD, with the four corners being roughly ($\Mr-5\log h$, $g-r$)=($-16.5$, 0.3), ($-19.5$, 0.3), ($-19.0$, 1.0), and ($-22.0$, 1.0). In searching for host galaxies of BNSM GW events, it would be beneficial to assign high observation priority to such galaxies in the search region and then expand the search to other galaxies (as the 95.4\% range goes over almost all the places in the CMD).

\subsection{Forecasts on DTD Constraints}
\label{sec:forecast}

The results in the previous subsection show the sensitivity of the galaxy property dependent occurrence probability of BNSM events to the DTD models. In what follows we show constraints on the DTD parameters from such host galaxy property distributions. With a given set of BNSM GW observations and host galaxies, we can apply such constraints to provide a quick estimate of the preferred DTD model, without inferring SFH of each host galaxy. Ultimately we would like to use the SFH of individual host galaxies to obtain the final DTD constraints, with all the information relevant to DTD accounted for. Therefore we also consider constraints from this most constraining case, denoted as `perGAL'.

The detection of BNSM events can be approximated as volume-limited, i.e. complete within a survey volume set by the sensitivity of GW observation. We perform forecasts on DTD constraints given the number $N_{\rm obs}$ of detections during a period of observations. We consider DTD models with $-2\leq n \leq 0$ and $-2.7\leq \log(\tm/{\rm Gyr})\leq 0.7$. For a given distribution of host galaxies from the truth model (i.e. the observation), the likelihood of DTD models are evaluated on a uniform grid in the $n$--$\log\tm$ plane, according to equation~(\ref{eq:DeltaL}) for cases with different galaxy properties or equation~(\ref{eq:DeltaL_perGAL}) for the perGAL case. 

\begin{figure*}
\includegraphics[width=\textwidth]{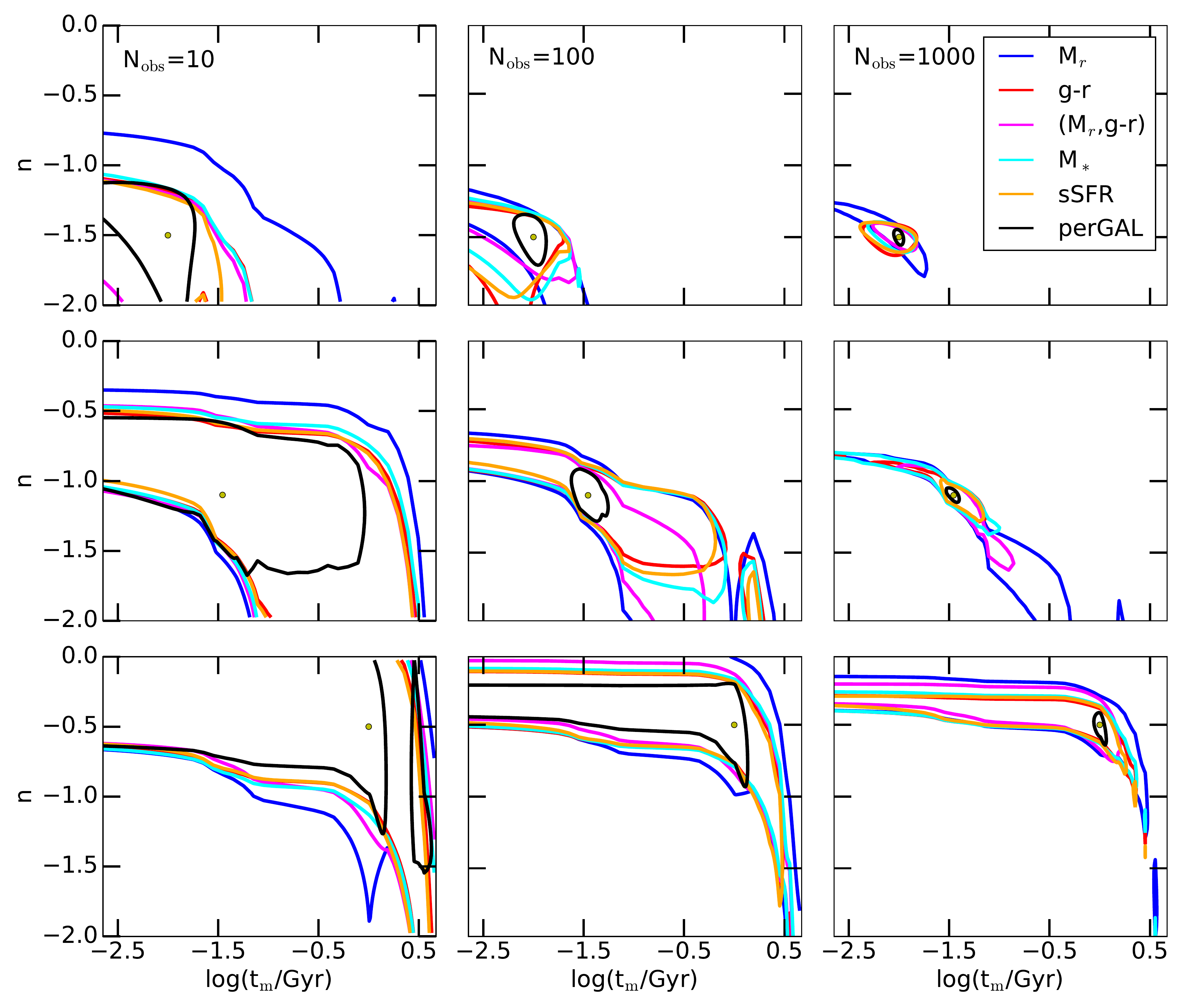}
\caption{
Constraints on DTD model parameters ($n$, $\tm$) based on distribution of properties of BNSM host galaxies. The calculation is done for a volume-limited sample of local galaxies. The top, middle, and bottom panels assume the truth model (denoted by the circle in each panel) to be the `Fast', `Canonical', and 'Slow' DTD model, respectively. The number of observed BNSM events is assumed to be 10, 100, and 1000 for the cases in the left, middle, and right panels. In each panel, constraints based on dependence on different galaxy properties are coded with different colours, with the contour marking the $1\sigma$ (68.3\%) confidence range for each case. The black contour shows the constraints from the SFH of individual galaxies. See text for detail. 
}
\label{fig:likecont}
\end{figure*}

Each row of Fig.~\ref{fig:likecont} shows the constraints on DTD parameters $n$ and $\tm$ for an assumed truth model (marked with the filled circle) and how the constraints improve as the number of observed BNSM events increases from 10 (left), to 100 (middle), and to 1000 (right). The top, middle, and bottom row corresponds to the case of truth model with `Fast', `Canonical', and 'Slow' DTD, respectively.

In each panel, the 68.3\% confidence contours from constraints related to different galaxy properties are shown.\footnote{The discontinuity of contours in a few panels are related to the treatment of thermally pulsating asymptotic giant branch (TP-AGB) stars in the stellar population synthesis model used to infer the SFH. See discussion in Section~\ref{sec:dis}.} 
As seen in previous work \citep[e.g.][]{Safarzadeh19a,Safarzadeh19b,Safarzadeh19c}, the constraints have an intrinsic degeneracy between the two DTD parameters. In fitting the observation, a DTD with smaller minimum delay time and flatter power law would be similar in likelihood to that with larger minimum delay time and steeper power law. Such a degeneracy direction is largely a manifestation of the overall decreasing star formation activity over the past $\sim$10 Gyr.

With 10 detections (left panels), the constraints based on various galaxy properties are quite loose. Those using luminosity distribution of host galaxies appear to be the least constrained, while the constraining powers from other properties (stellar mass, colour, colour+magnitude, and sSFR) are all similar. Using SFH of individual host galaxies (the perGAL case) improves the constraints, while still loose. Nevertheless, with 10 detections, we would be able to differentiate substantially different DTD models. For example, with `Canonical' DTD as the truth model, the `Fast' DTD with $n=-1.5$ and $\tm=0.01$ Gyr can be ruled out at $2.1\sigma$ confidence level. Similarly, with `Fast' DTD as the truth model, the `Slow' DTD model can be excluded at $3.6\sigma$ confidence level.

With 100 detections (middle panels), the constraints with various galaxy properties all improve, and those with luminosity distribution are still the least constrained. The perGAL constraints have been improved a lot, with substantially shrunk contours (black) with respect to the case of 10 detections and to those with galaxy properties, and the shape of contours becomes close to ellipse (except for the `Slow' truth model case). With 1000 detections, the perGAL method provides tight constraint on both parameters, while those from all other methods appear to be mostly thin bands following the degeneracy direction (except for the case with the `Fast' truth model constrained based on other than the luminosity dependence).

\begin{figure*}
\includegraphics[width=\textwidth]{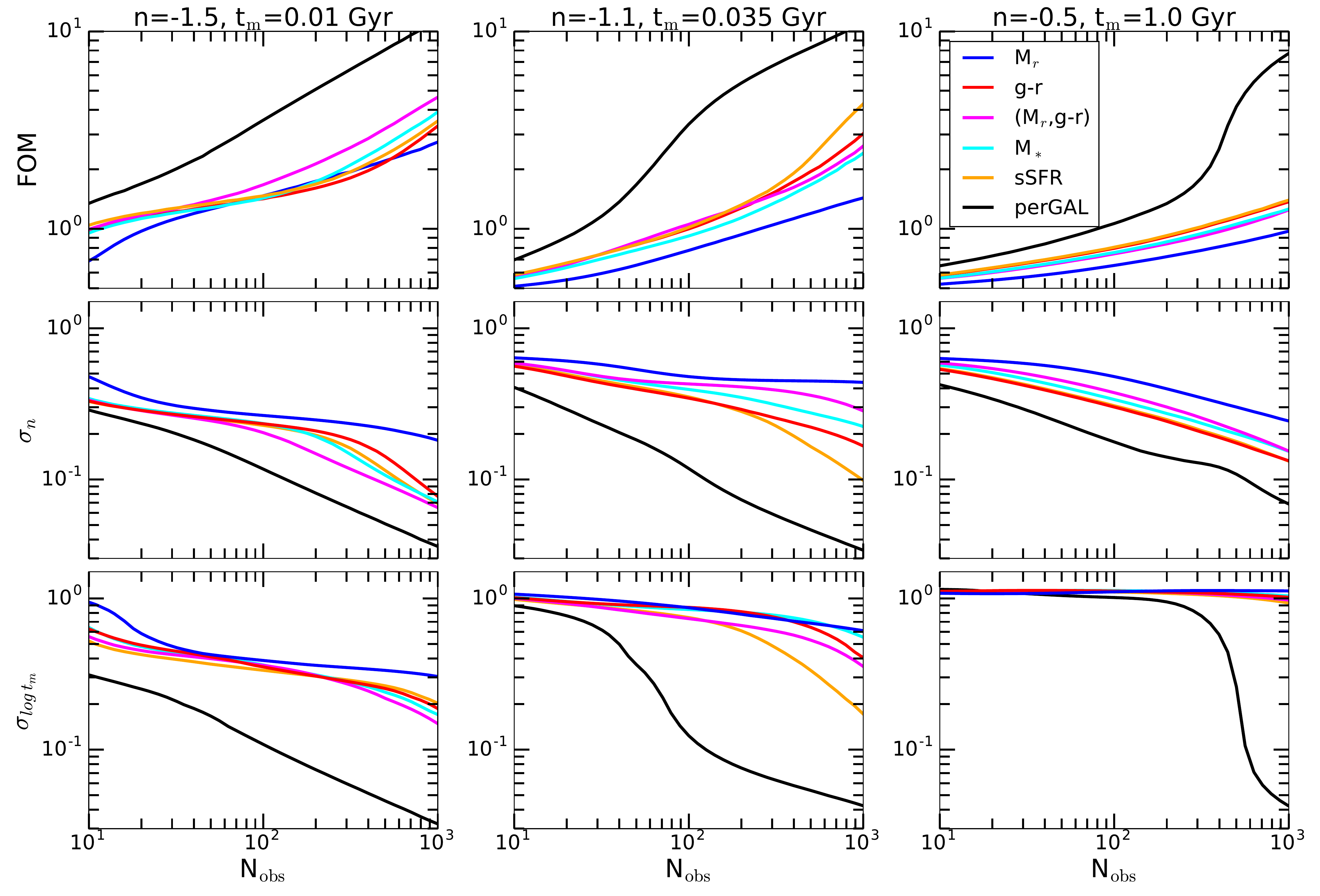}
\caption{
Figure of merit (FOM) and uncertainties in DTD parameter constraints as a function of the number of BNSM observations. Panels from left to right correspond to the three truth models (`Fast', `Canonical', and 'Slow' model, with parameters shown on the top). Top panels show the FOM curves from the dependence of BNSM occurrence on different galaxy properties (same colour code as in Fig.~\ref{fig:likecont}). The FOM is defined as the inverse square root of the area of the 68.3\% confidence contour in the $n$--$\log\tm$ plane. Middle and bottom panels show the corresponding 1$\sigma$ uncertainties in the DTD parameters $n$ and $\log t_m$, respectively.
}
\label{fig:FOM}
\end{figure*}

To quantify the constraints from different methods and the improvement with the number of observations, we compute a figure of merit (FOM; e.g. \citealt{DETF06}) in constraining $n$ and $\log\tm$. We define the FOM to be the inverse square root of the area of the 68.3\% confidence contour, which can be regarded as being proportional to the reciprocal of an average uncertainty in the $n$--$\log\tm$ constraints. 

The top panels of Fig.~\ref{fig:FOM} show the values of FOM from different methods of constraints and their dependence on the number of detections. Given that the log-likelihood is proportional to $\Nobs$ [equations~(\ref{eq:DeltaL}) and (\ref{eq:DeltaL_perGAL})], a two-dimensional  (2D) Gaussian likelihood approximation around the maximum would predict that the FOM scales as $\sqrt{\Nobs}$. This appears to be the case for sufficiently large $\Nobs$. At small $\Nobs$, since the likelihood surface is not well described by a 2D Gaussian and the 1$\sigma$ contours in most cases are not closed (Fig.~\ref{fig:likecont}) as a result of reaching the boundary of priors imposed in the calculation, the increase of the FOM deviates from the $\sqrt{\Nobs}$ scaling. The FOM of the constraints with the `Slow' truth model has the slowest transition to the $\sqrt{\Nobs}$ scaling regime, at $\Nobs\gtrsim 500$. In the high $\Nobs$ regime, the FOM from the perGAL method is typically a factor of more than three higher than any of the other methods. 

From the marginalised distribution, we obtain the 1$\sigma$ uncertainty in each DTD parameter from  constraints with each method, as shown in the middle and bottom panel of Fig.~\ref{fig:FOM}. As $\Nobs$ increases, we expect the uncertainty to decrease as $1/\sqrt{\Nobs}$, given how the likelihood depends on $\Nobs$. 

For most of the methods, this scaling relation shows up at sufficiently large $\Nobs$. However, with the `Slow' truth model, the constraint on the parameter $t_m$ does not improve substantially even with 1,000 detections (bottom-right panel).

With the perGAL method, such a scaling relation works well except for $\log \tm$ constraints at $\Nobs\lesssim 100$ (500) for the case of `Canonical' (`Slow') truth model.  For those two models in the low $\Nobs$ regime, the constraints on $\log \tm$ are loose, which is echoed in the corresponding FOM values (top panels) and also evident in Fig.~\ref{fig:likecont}. 

While $\mathcal{O}(10)$ BNSM detections from GW observations are able to differentiate substantially different DTD models (e.g. `Fast' versus `Slow' model),  
precise constraints on DTD parameters require more detections. With the most constraining method (perGAL), if the DTD is close to the `Slow' model, constraining the model is not easy -- about 600 BNSM detections are needed to reach $\sim$10\% precision on the constraints of $n$ and $\log \tm$. For DTD close to the other two models, we only need about $160$ detections to reach 10\% precision on the constraints of both parameters.

\section{Discussion}
\label{sec:dis}

We investigate the distribution of properties of BNSM host galaxy by combining a catalogue of local SDSS galaxies and a parameterised DTD model. Relevant studies have been performed with simulated galaxy catalogues and variations of DTD models, and we find broad agreements for relevant results. For example, \citet{Artale19} and \citet{Artale20} study the correlation of BNSM rate and galaxy properties by applying a population synthesis DTD model to galaxies in hydrodynamic simulations. \citet{Adhikari20} show the distribution of BNSM host galaxies using galaxies from Universe Machine simulations. 

The forecasts on DTD parameter constraints have been carried out using stellar mass dependent analytic SFH \citep{Safarzadeh19a} or the SFH of galaxies from simulation \citep{Adhikari20}. \citet{Safarzadeh19c} use  SFH of individual galaxies inferred from galaxy photometry to investigate the DTD model constraints, and our results are in agreement with theirs. While a large number of realisations of observation are used in \citet{Safarzadeh19c} to make the forecast, no realisation is performed in our investigation by adopting the formalism we develop. Effectively our method can be regarded as performing a mean realisation.  While realisations have the advantages to account for the sample variance effect (e.g. in shifting the central values), for the purpose of model forecast our formalism works well and is more efficient.

In our study, we focus on the distribution of BNSM events with galaxy property, not the absolute rate. Given the number of observations and the observation period, the absolute rate can be estimated. To make the corresponding forecast within our formalism, we note that the absolute rate is encoded in the normalisation constant $C$ in equation~(\ref{eq:rate}) and we just need to keep the $\Nmod$ and $\Nobs$ terms in equation~(\ref{eq:fullLike}).

When making the forecast, we implicitly neglect any uncertainty in the SFH of galaxies. In this work, the SFH is inferred using the BC03 stellar population synthesis model. If instead we use that from \citeauthor{Maraston05} (\citeyear{Maraston05}; M05), the details in our results would change. The M05 model includes TP-AGB stars, which makes the stellar population with age around 1 Gyr more luminous and leads to lower amount of stellar mass needed in populations of this age. The overall effect is a shallower decay of SFH (see fig.15 and fig.16 of \citealt{Tojeiro09}). The discontinuity of some contours in our Fig.~\ref{fig:likecont} at $\tm \approx$ 1 Gyr is likely caused by the higher stellar mass (thus higher BNSM rate) in populations of such ages inferred using the BC03 model that neglects the TP-AGB contribution. Also different ways of modelling the dust effect can lead to differences in the inferred SFH, which mainly affects populations with age younger than 0.1 Gyr (fig.20 of \citealt{Tojeiro09}).

In principle, the systematic uncertainties in SFH modelling and inference should be incorporated into DTD model constraints, especially when model parameters start to be tightly constrained by BNSM observations. Also
at such a stage DTD models more sophisticated than the simple two-parameter model can be tested (such as those including the effect of metallicity, e.g. \citealt{Artale20}).

\section{Summary and Conclusion}
\label{sec:sum}

We combine a catalogue of local SDSS galaxies with inferred SFH and a parameterised BNSM DTD model to investigate the dependence of BNSM rate on an array of galaxy properties, including galaxy colour ($g-r$), luminosity ($\Mr$), stellar mass, and sSFR. We introduce a formalism to efficiently make forecast on using BNSM detections from GW observations to constrain DTD models, and we then predict the constraints based on galaxy property dependent BNSM occurrence distribution and based on SFH of individual host galaxies.

Compared to the intrinsic property distribution of galaxies, the distribution of BNSM host galaxies is skewed toward galaxies with redder colour, lower sSFR, higher luminosity, and higher stellar mass, largely reflecting the tendency of higher BNSM rates in more massive galaxies. Based on three DTD models, corresponding to fast, canonical, and slow merger scenarios, the host galaxies of BNSM events are likely concentrated in a broad band across the galaxy CMD, ranging from ($\Mr-5\log h$, $g-r$)=($-18.0\pm 1.5$, 0.3) to ($-20.5 \pm 1.5$, 1.0), which can be assigned high priorities for searching for EM counterparts.

The efficient forecast formalism introduced in this work is in a form of relative entropy of two distributions, which can have wide applications in constraining distributions in various astrophysical situations. In particular, it can be applied to study DTD of other transient events associated with galaxy SFH, such as short GRBs \citep[e.g.][]{Zheng07,Leibler10,Behroozi2014},  supernova Ia \citep[e.g.][]{Aubourg08,Maoz12}, and potentially neutron star -- black hole mergers and black hole -- black hole mergers (as long as black holes are of stellar origin to be related to SFH and host galaxies can be identified). The formalism can also be extended to higher redshifts for such studies, as long as the SFH of individual galaxies is available for a galaxy sample at the redshift of interest.

In this work, we consider power-law DTD models with a minimum delay time, represented by the power-law index $n$ and the cutoff time scale $\tm$. Constraints on the DTD model can be obtained based on property distribution of BNSM host galaxies, without inferring their SFH. The constraints depend on how tight the correlation is between the galaxy property and the SFH. As with previous study \citep[e.g.][]{Safarzadeh19a,Artale20,Adhikari20}, we find that galaxy colour, stellar mass, and sSFR are good predictors of BNSM rate, as well as the joint colour and luminosity information. Using the dependence on host galaxy luminosity alone usually produces the weakest constraints, with FOM in some cases reduced by about 50\%, where the FOM is defined as the inverse square root of the area of the 1$\sigma$ contour in the $n$--$\log \tm$ plane.  

Given a set of BNSM detections, the tightest constraints on DTD models are obtained by using the individual SFH of host galaxies, with the FOM enhanced by a factor of three or more compared to the galaxy property based constraints. In line with \citet{Adhikari20}, we find that $\mathcal{O}(10)$ detections would be able to tell apart substantially different DTD models. For precision DTD constraints, a much larger sample of BNSM events with identified host galaxies are necessary, e.g. a few hundred events for $\sim$10\% constraints on either $n$ 
or $\log\tm$, in broad agreement with the result in \citet{Safarzadeh19c}. If we adopt $\sim$160 detections as the requirement (Section~\ref{sec:forecast}) and assume the sensitivity of aLIGO O4 run (corresponding to a BNSM detection horizon of $\sim$160--190 Mpc; \citealt{Abbott13}) and the estimated local BNSM rate of 250--2810 ${\rm Gpc^{-3}yr^{-1}}$ \citep{Abbott20}, such a precision can be achieved in $\sim$2--40 years. 

\section*{Acknowledgements}

KSM acknowledges the support by a fellowship from the Willard L. and Ruth P. Eccles Foundation. The support and resources from the Center for High Performance Computing at the University of Utah are gratefully acknowledged.  E.R.-R. is grateful for support from the  The Danish National Research Foundation (DNRF132) and NSF grants (AST-161588, AST-1911206 and AST-1852393).

\section*{Data Availability}
No new data were generated or analysed in support of this research.

\bibliographystyle{mnras}
\bibliography{gw}
\bsp    
\label{lastpage}
\end{document}